\documentclass[preprint,12pt,authoryear,nopreprintline]{elsarticle}

\usepackage{amsmath,amssymb,amsthm}
\usepackage{booktabs}
\usepackage{graphicx}
\usepackage{array}
\usepackage[hidelinks]{hyperref}
\usepackage[margin=1in]{geometry}
\usepackage{setspace}
\providecommand{\doi}{}\renewcommand{\doi}[1]{\url{https://doi.org/#1}}

\journal{International Journal of Forecasting}

\newtheorem{proposition}{Proposition}
\newtheorem{corollary}{Corollary}
\theoremstyle{definition}
\newtheorem{definition}{Definition}
\newtheorem{remark}{Remark}

\newcommand{\ip}[2]{\langle #1,#2\rangle_t}
\newcommand{\nrm}[1]{\lVert #1\rVert_t}
\newcommand{\one}{\mathbf{1}}
\newcommand{\Kpool}{K^{\mathrm{pool}}}
\newcommand{\Cbar}{\bar C}
\newcommand{\Kbar}{\bar K}
\newcommand{\gbar}{\bar\gamma}
\newcommand{\Ctil}{\widetilde C}
\newcommand{\Ktil}{\widetilde K}
\newcommand{\gtil}{\widetilde\gamma}
\newcommand{\Rrel}{\mathcal R}
\newcommand{\Dsf}{\Delta^{\mathrm{sf}}}
\newcommand{\Dsc}{\Delta^{\mathrm{scale}}}

\begin{document}
\sloppy

\begin{frontmatter}

\title{Target alignment, dilution and forecast selection when cross-sectional forecasts share a common target}

\author{Masoud Soleimani}
\ead{m.soleimani@ieee.org}
\affiliation{organization={Department of Information Engineering, University of Pisa},
             city={Pisa},
             country={Italy}}

\begin{abstract}
Forecasters often score the same units per date against one standardized realized outcome. We show that every standardized forecast splits exactly into a component aligned with this common target and a component uncorrelated with it. Three consequences follow: forecast-error correlation largely mirrors forecast correlation and is therefore a poor measure of diversity; an equally weighted combination beats a no-information forecast only when average alignment is large relative to the combination's dispersion; and the gain from adding a forecaster separates into genuine improvement and mere dilution, which equal-weight admission can mistakenly reward. We develop a cautious selection rule, study it in simulations, and apply it to language-model forecasts of US equity rankings and mechanical signals ranking exchange-traded funds. Selection removes most dilution losses, but no combination beats the no-information forecast.
\end{abstract}


\end{frontmatter}

\section{Introduction}\label{sec:intro}

Forecast combination ranks among the most reliable routes to improved accuracy \citep{bates1969,clemen1989,timmermann2006,wang2023review}, and simple averages are notoriously difficult to beat \citep{stockwatson2004,genre2013,claeskens2016}. The conventional rationale rests on diversification: forecasts whose errors behave differently offset one another's mistakes \citep{batchelor1995,lichtendahl2020}. Accordingly, the value of adding a forecaster is typically gauged by how weakly its errors correlate with those already included \citep{krogh1995,brown2005,magnus2023,kim2026}.

This paper investigates a setting where that rule of thumb turns ambiguous. On each date a forecaster scores a large number of units (assets, products, regions), the scores are standardized across units, and accuracy is measured against the standardized realized outcome. Examples include rankings of expected relative returns \citep{jegadeeshtitman1993,asness2013}, relative demand across stores, and relative growth across regions. Every forecast on a given date is evaluated against the \emph{same} standardized target. Consequently, the correlation of forecast deviations from that target confounds how strongly forecasts move together with how strongly each moves with the target.

The framework rests on a single exact decomposition. Each standardized forecast equals the sum of its projection onto the standardized target, whose coefficient is the forecaster's \emph{target alignment}, and a \emph{target-orthogonal component}. The forecast correlation matrix therefore equals the outer product of the alignments plus the covariance of the orthogonal components. The paper makes three contributions.
\begin{enumerate}
\item \textbf{Geometry of common-target combination.} We derive exact implications of the split for relative-score risk. Deviation correlation equals a translation of forecast correlation plus an alignment term. A closed-form condition characterizes when a combination beats the no-information forecast, and an in-sample bound yields the best risk achievable by any linear combination. After time aggregation, the average within-date orthogonal covariance and the pooled Schur complement differ by the time variation of alignment.
\item \textbf{Incremental risk of equal-weight admission.} We derive exact marginal and batch conditions for adding forecasts to an equal-weight pool, and decompose the incremental risk into a scale-free component and a scale-mismatch component. A dilution bound reveals why equal-weight admission into a large pool is unresponsive to aligned candidates. It further demonstrates why the unscaled rule can admit forecasts that merely shrink the composite toward zero. On this foundation we develop a three-way admission rule (admit, reject, undecided) with simultaneous heteroskedasticity- and autocorrelation-consistent (HAC) bounds inside nested rolling-origin validation.
\item \textbf{Evidence on what the geometry implies in practice.} Seven simulation designs encompass: when combination can help at all (a phase diagram); which history statistics rank candidates by their future incremental risk (criterion validity); how selection, subset averaging and regression-type combination trade off across history lengths and instability mechanisms; and behavior under misspecification of the generating geometry. Two empirical panels, a date-shifted placebo, a mechanical control panel and a planted-signal positive control illustrate the framework on real data and calibrate the power behind its null findings.
\end{enumerate}

Although the empirical findings pertain to the two panels examined, they replicate across both: forecast correlation accounts for nearly all of the common-target deviation correlation; equal-weight admission rewards dilution; selection and weighting eliminate most of the dilution loss carried by the full equal-weight pool, yet fail to improve on the no-information forecast; and neither panel displays detectable alignment. A date-shifted placebo, together with planted-signal experiments, establishes what these null results can and cannot support.

We do not assert that one dependence measure is universally superior, that equal-weight admission is optimal, that the three-way rule controls error rates along an adaptive selection path, or that large language models (LLMs) cannot forecast returns. Table~\ref{tab:status} summarizes the status of each result.

\subsection{Relation to existing work and what is new}\label{sec:new}

\emph{Ambiguity and bias--variance--covariance decompositions} decompose ensemble loss into individual losses and disagreement \citep{krogh1995,ueda1996,brown2005,wood2023}. We derive their relative-score analogue (Proposition~\ref{prop:ambiguity}). The novel element is the decomposition of dependence itself into alignment and orthogonal parts, which reveals that disagreement is target-free while deviation correlation is not.

\emph{Error-covariance and encompassing approaches} weight or test forecasts through their error covariance \citep{bates1969,granger1984,chong1986,harvey1998}. With a common standardized target, the error covariance becomes an explicit function of forecast correlation and alignment (Proposition~\ref{prop:rhoe}). The scale-free component of incremental risk constitutes an equal-weight, pool-level analogue of an incremental-$R^2$ encompassing statistic, batch admission is the corresponding analogue of multiple encompassing \citep{harveynewbold2000}, and the scale-mismatch component represents the risk of a miscalibrated composite in the sense of \citet{mincer1969}.

\emph{Factor and idiosyncratic decompositions} distinguish common from forecaster-specific components of forecasts or errors and exploit the latter for combination \citep{leelee2026,leeseregina2026}. In that literature, the split constitutes a statistical model estimated from the data. Ours is an exact, model-free projection on the realized target, and the orthogonal component is defined relative to the target rather than to the other forecasts.

\emph{Model-importance measures} evaluate the loss of an ensemble with and without a component \citep{budescu2015,kim2026}. Under relative-score loss our incremental risk constitutes such a measure in closed form. The novel element is its exact split into scale-free and scale-mismatch parts, and the dilution bound demonstrating that equal-weight contributions diminish with pool size.

\emph{Selection before averaging}, by trimming, clustering, performance screening or model confidence sets, can outperform the simple average \citep{aiolfi2006,hansen2011mcs,burgi2017,matsypura2018,dieboldshin2019}, while estimated weights suffer from estimation error \citep{stockwatson2004,smithwallis2009,claeskens2016} and may involve negative weights for similar forecasters \citep{radchenko2023}; constraining the weights or shrinking them toward equality mitigates this \citep{conflitti2015,roccazzella2022}. We compare three-way admission, the partially egalitarian LASSO (peLASSO) of \citet{dieboldshin2019} and regression-type combination within one geometry. We report where each wins, including cases where selection is dominated.

\begin{table}[t]\centering\small
\caption{Position relative to adjacent approaches. ``Scale separated'' means the method distinguishes a change in the composite's correlation with the target from a change in its scale.}\label{tab:position}
\resizebox{\textwidth}{!}{%
\begin{tabular}{@{}lllll@{}}
\toprule
Approach & Object & Weights & Scale separated & Status \\
\midrule
Ambiguity / bias--variance--covariance & disagreement, error covariance & convex & no & identity \\
Forecast encompassing & incremental predictive content of one forecast & regression & implicitly (by regression) & test \\
Regression combination and shrinkage & error covariance and means & unrestricted or penalized & implicitly & estimator \\
Factor / idiosyncratic decompositions & estimated common and specific parts & penalized & no & estimated model \\
Model-importance measures & ensemble loss with and without a member & equal or given & no & estimated measure \\
Subset averaging (e.g.\ peLASSO) & support of a penalized fit & equal on support & no & estimator \\
\textbf{This paper} & \textbf{alignment and target-orthogonal dependence} & \textbf{any, incl.\ equal and rescaled} & \textbf{yes, exactly} & \textbf{identity + decision rule} \\
\bottomrule
\end{tabular}}
\end{table}

\begin{table}[t]\centering\small
\caption{Status of the main results. ``Identity'' results hold exactly for any sample and any weights; ``procedure'' results depend on estimation and on assumptions stated in the text.}\label{tab:status}
\resizebox{\textwidth}{!}{%
\begin{tabular}{@{}llp{9.2cm}@{}}
\toprule
Result & Status & Scope \\
\midrule
Propositions~\ref{prop:decomp}--\ref{prop:rhoe}, \ref{prop:agg}, \ref{prop:admission}--\ref{prop:dsplit}; Corollary~\ref{cor:null} & Algebraic identity & Exact for standardized forecasts on a common support; no distributional assumptions \\
Corollary~\ref{cor:attain} (attainable risk) & Identity, in-sample & A bound for the sample moments; not an operationally attainable out-of-sample risk \\
Proposition~\ref{prop:shrink} (joint shrinkage) & Estimator property & PSD preservation for valid inputs; not a repair for pairwise deletion \\
Three-way rule & Procedure & Simultaneous per-step bands conditional on the incumbent pool, asymptotically valid under HAC consistency; no guarantee along the adaptive greedy path \\
Joint zero-alignment tests & Procedure & Bootstrap Wald and max-$|t|$ size and power assessed by simulation; asymptotic Wald oversized \\
Decision error rates, criterion validity, method rankings & Simulation finding & Specific to the designs in Section~\ref{sec:simulations} \\
Geometry, nulls, admission and positive-control results & Application finding & Specific to the two panels in Sections~\ref{sec:empirical} and~\ref{sec:panelB}: a retrospective fixed-panel language-model experiment and a cross-asset exchange-traded fund (ETF) panel \\
Placebo (date-shifted target) & Application finding & A randomization reference for the alignment statistics; it requires no distributional assumption \\
\bottomrule
\end{tabular}}
\end{table}

\section{Estimand, geometry and notation}\label{sec:geometry}

\subsection{Common support, standardization and the estimand}

At date $t$, let $r_t$ denote the realized cross-sectional target and $x_{it}$ the raw score vector of forecaster $i\in\{1,\dots,N\}$. All inner products at $t$ are calculated on the common support $\mathcal M_t^\ast$ of units possessing a valid target and valid forecasts from every forecaster, with $M_t=|\mathcal M_t^\ast|$. Pairwise deletion is employed solely as a robustness check, since it destroys the Gram-matrix structure and can produce indefinite matrices that would require a nearest-correlation repair \citep{higham2002}. Define $\ip{a}{b}=M_t^{-1}a'b$ and standardize using the same $1/M_t$ normalization:
\begin{equation}
s_{it}=\frac{x_{it}-\bar x_{it}\one}{\hat\sigma_{x_i,t}},\qquad y_t=\frac{r_t-\bar r_t\one}{\hat\sigma_{r,t}},\qquad \nrm{s_{it}}=\nrm{y_t}=1.
\end{equation}
Dates whose target is degenerate ($\hat\sigma_{r,t}\le\varepsilon_r$) are excluded. A forecaster with $\hat\sigma_{x_i,t}\le\varepsilon_x$ is unavailable at $t$, which eliminates that date in the fixed-membership analysis. Missing outputs are not imputed.

The loss of a combination $s_{w,t}=\sum_iw_is_{it}$ is $\nrm{s_{w,t}-y_t}^2$. It evaluates relative forecasts after each date's location and scale of the target have been removed, which is appropriate for tasks whose outputs constitute rankings or relative scores with arbitrary level and dispersion. For a unit-norm composite, the loss equals $2(1-\operatorname{corr}_t)$, twice one minus the cross-sectional correlation. For an unscaled composite, it additionally penalizes the composite's norm (Proposition~\ref{prop:scale}). Rank correlations and top-minus-bottom return spreads are reported as diagnostics. The loss does not constitute a portfolio utility, and statistical and economic rankings of forecasts need not coincide \citep{leitchtanner1991,gneiting2011}. The zero forecast $w=0$ has loss exactly one. We term it the \emph{no-information forecast}: it serves as the natural benchmark \emph{under this estimand}, and ``beating'' it signifies improving the standardized relative-score risk, not generating economic value.

\subsection{Target alignment and the target-orthogonal component}

\begin{definition}
Target alignment is $\gamma_{it}=\ip{s_{it}}{y_t}$, a correlation since both vectors possess unit norm. The target-orthogonal component is $u_{it}=s_{it}-\gamma_{it}y_t$. The forecast correlation matrix is $C_t=[\ip{s_{it}}{s_{jt}}]$ with entries $\rho^s_{ij,t}$, and the target-orthogonal covariance is $K_t=[\ip{u_{it}}{u_{jt}}]$.
\end{definition}

The orthogonal component constitutes an exact geometric residual. It is not a latent signal, an independent economic factor, or a source of predictive value. Table~\ref{tab:notation} assembles the notation.

\begin{table}[t]\centering\small
\caption{Notation. Date-level objects carry $t$; bars denote averages over a window with weights $a_t$ (uniform here); tildes denote shrinkage estimates.}\label{tab:notation}
\resizebox{\textwidth}{!}{%
\begin{tabular}{@{}lll@{}}
\toprule
Object & Definition & Measures \\
\midrule
$\gamma_{it}$, $\gamma_t$ & $\ip{s_{it}}{y_t}$ & target alignment (correlation with the standardized target) \\
$C_t$, $\rho^s_{ij,t}$ & $\ip{s_{it}}{s_{jt}}$ & forecast correlation (target-free) \\
$K_t$ & $\ip{u_{it}}{u_{jt}}=C_t-\gamma_t\gamma_t'$ & target-orthogonal covariance \\
$\rho^e_{ij,t}$ & correlation of $s_{it}-y_t$ and $s_{jt}-y_t$ & common-target deviation correlation \\
$\Cbar,\ \gbar,\ \Kbar,\ G$ & $\sum_ta_tC_t,\ \sum_ta_t\gamma_t,\ \sum_ta_tK_t,\ \sum_ta_t\gamma_t\gamma_t'$ & time-averaged objects \\
$\Kpool$ & $\Cbar-\gbar\gbar'=\Kbar+\operatorname{Cov}_a(\gamma_t)$ & pooled orthogonal covariance \\
$\Rrel(w)$ & $\sum_ta_t\nrm{s_{w,t}-y_t}^2$ & relative-score risk \\
$g_w,\ q_w,\ c_w,\ \rho_w$ & $w'\gbar,\ w'\Cbar w,\ g_w/q_w,\ g_w/\sqrt{q_w}$ & mean alignment, squared norm, risk-minimizing scale, pooled correlation \\
$V_P$, $\Delta_{A|P}$ & risk of the equal-weight pool $P$; $V_{P\cup A}-V_P$ & incremental relative-score risk \\
$\Dsf_{A|P},\ \Dsc_{A|P}$ & $\rho^2_P-\rho^2_{P\cup A}$; $\Delta_{A|P}-\Dsf_{A|P}$ & scale-free and scale-mismatch components \\
$\delta$ & minimum practically relevant improvement & threshold of the three-way rule \\
\bottomrule
\end{tabular}}
\end{table}

\begin{proposition}[Target-orthogonal decomposition]\label{prop:decomp}
$\ip{u_{it}}{y_t}=0$, $\nrm{u_{it}}^2=1-\gamma_{it}^2$, and $C_t=\gamma_t\gamma_t'+K_t$ with $K_t\succeq0$. Furthermore $\operatorname{rank}(C_t)\le\min(N,M_t-1)$ and $\operatorname{rank}(K_t)\le\min(N,M_t-2)$.
\end{proposition}

\section{Relative-score risk}\label{sec:risk}

\begin{proposition}[Risk decomposition]\label{prop:risk}
For any $w\in\mathbb R^N$ (affine, convex, rescaled or zero), $\nrm{s_{w,t}-y_t}^2=(1-w'\gamma_t)^2+w'K_tw=1-2w'\gamma_t+w'C_tw$. Consequently $\Rrel(w)=1-2g_w+q_w$.
\end{proposition}

We refer to $(1-w'\gamma_t)^2$ as the \emph{alignment term} and $w'K_tw$ as the \emph{target-orthogonal term}.

\begin{corollary}[Improving on the no-information forecast]\label{cor:null}
$\Rrel(w)<1$ if and only if $g_w>\tfrac12q_w$. For equal weights across $N$ forecasters with mean alignment $\gbar_{\mathrm{EW}}$ and mean off-diagonal correlation $\bar\rho$,
\begin{equation}\label{eq:ewnull}
\Rrel(\one/N)<1\iff \gbar_{\mathrm{EW}}>\tfrac12\bigl[\bar\rho+(1-\bar\rho)/N\bigr],
\end{equation}
and the margin $m=\gbar_{\mathrm{EW}}-\tfrac12\one'\Cbar\one/N^2$ satisfies $\Rrel(\one/N)-1=-2m$.
\end{corollary}

Averaging standardized forecasts draws the composite toward zero. Unless mean alignment exceeds half the composite's squared norm, the no-information forecast exhibits lower risk, regardless of how many forecasters are averaged.

\begin{corollary}[In-sample attainable risk]\label{cor:attain}
$\min_{v}\{1-2v'\gbar+v'\Cbar v\}=1-\gbar'\Cbar^{+}\gbar\in[0,1]$, achieved at $v^\ast=\Cbar^{+}\gbar$. The quantity $\gbar'\Cbar^{+}\gbar$ represents the pooled squared multiple correlation of the standardized target on the forecasts in the window.
\end{corollary}

\begin{proposition}[Scale split]\label{prop:scale}
For $q_w>0$, $\min_c\Rrel(cw)=\Rrel(c_ww)=1-\rho_w^2$ and
\begin{equation}\label{eq:scale}
\Rrel(w)=(1-\rho_w^2)+q_w(1-c_w)^2.
\end{equation}
\end{proposition}

The first term constitutes the risk of the optimal rescaling of $w$ and depends solely on the pooled correlation $\rho_w$. The second is the \emph{scale-mismatch penalty} of the unscaled composite. When alignment is weak, $c_w\approx0$ and the penalty approximates $q_w$, so ranking unscaled combinations by risk largely amounts to ranking them by their norm.

\section{Disagreement and common-target deviation correlation}\label{sec:deviation}

\begin{proposition}[Ambiguity]\label{prop:ambiguity}
For convex $w$, $\mathcal A_t(w)=\sum_iw_i\nrm{s_{it}-s_{w,t}}^2=1-w'C_tw$, and with $L_{it}=\nrm{s_{it}-y_t}^2=2(1-\gamma_{it})$, $\sum_iw_iL_{it}=\nrm{s_{w,t}-y_t}^2+\mathcal A_t(w)$.
\end{proposition}

Disagreement depends on $C_t$ alone; alignment enters through individual losses. Disagreement therefore does not on its own imply a forecasting benefit.

\begin{proposition}[Common-target deviation correlation]\label{prop:rhoe}
With $e_{it}=s_{it}-y_t$, $\ip{e_{it}}{e_{jt}}=1+\rho^s_{ij,t}-\gamma_{it}-\gamma_{jt}$, $\nrm{e_{it}}^2=2(1-\gamma_{it})$, and for $\gamma_{it},\gamma_{jt}<1$
\begin{equation}\label{eq:rhoe}
\rho^e_{ij,t}=\frac{1+\rho^s_{ij,t}-\gamma_{it}-\gamma_{jt}}{2\sqrt{(1-\gamma_{it})(1-\gamma_{jt})}}.
\end{equation}
At zero alignment $\rho^e_{ij,t}=(1+\rho^s_{ij,t})/2$. For small alignments $\rho^e_{ij,t}-(1+\rho^s_{ij,t})/2=-\tfrac14(\gamma_{it}+\gamma_{jt})(1-\rho^s_{ij,t})+O(\gamma^2)$.
\end{proposition}

The claim that ``$\rho^e$ is a translation of $\rho^s$'' is not merely a restatement of the error covariance. What matters is that, with a common standardized target, the translation is fixed ($[-1,1]\mapsto[0,1]$ at zero alignment, halving dispersion). Its departure from $(1+\rho^s)/2$ is driven by alignment alone. The statistic is therefore not, in general, a clean measure of forecast-output diversity. Nor is it useless: because it embeds alignment, it can rank candidates better than forecast correlation (Section~\ref{sec:simulations}). The \emph{equal-weight variance-equivalent ensemble size} $N_{\mathrm{eff}}=N/[1+(N-1)\bar\rho]$ is a variance ratio, not a count of independent models, although under exchangeable dependence it coincides with the equivalent number of independent experts of \citet{clemenwinkler1985}. Computed from $\bar\rho^e$ it inherits the translation.

\section{Time aggregation and joint shrinkage}\label{sec:aggregation}

\begin{proposition}[Aggregation]\label{prop:agg}
$\Cbar=G+\Kbar$ and $\Kpool=\Cbar-\gbar\gbar'=\Kbar+\operatorname{Cov}_a(\gamma_t)$. Consequently $\Rrel(w)=\sum_ta_t(1-w'\gamma_t)^2+w'\Kbar w=(1-g_w)^2+w'\Kpool w$.
\end{proposition}

The two exact decompositions attribute the same risk in different ways. Time variation of $w'\gamma_t$ is charged to the alignment term within dates and to the orthogonal term once pooled. Risks, not daily ratios, are aggregated.

For dependence estimation we shrink the joint matrix $\widehat R=\bigl(\begin{smallmatrix}\Cbar&\gbar\\ \gbar'&1\end{smallmatrix}\bigr)$ toward a structured target $R_0$. The target has an equicorrelated block with the history mean correlation $\rho^\star$, as in constant-correlation shrinkage \citep{ledoitwolf2004jpm}, and a common alignment $g^\star$, projected onto $(g^\star)^2\le[1+(N-1)\rho^\star]/N$ if necessary. The estimate is $\widetilde R=(1-\lambda)\widehat R+\lambda R_0$, with $\lambda$ chosen by Frobenius fit from a fit window to a validation window, and $\Ktil=\Ctil-\gtil\gtil'$.

\begin{proposition}[Structure preservation]\label{prop:shrink}
If $\widehat R\succeq0$ and $R_0\succeq0$ have unit diagonals, then $\widetilde R$ is a valid correlation matrix and $\Ktil\succeq0$ for all $\lambda\in[0,1]$.
\end{proposition}

Joint shrinkage keeps $C$, $\gamma$ and $K$ mutually compatible \citep[cf.][]{ledoitwolf2004}. It serves the weighted benchmarks and diagnostics; admission decisions rely on realized losses.

\section{Incremental risk of equal-weight admission}\label{sec:admission}

For an equal-weight pool $P$ of size $n$ let $e_{P,t}=n^{-1}\sum_{i\in P}s_{it}-y_t$, $V_P=\sum_ta_t\nrm{e_{P,t}}^2$, and for a batch $A$ of size $q$ let $C^e_{PA}=\sum_ta_t\ip{e_{P,t}}{e_{A,t}}$ be the pool--batch \emph{deviation} covariance.

\begin{proposition}[Admission algebra]\label{prop:admission}
$V_{P\cup A}=[n^2V_P+q^2V_A+2nqC^e_{PA}]/(n+q)^2$ and $\Delta_{A|P}=V_{P\cup A}-V_P$. For an individual candidate $k$, $\Delta_{k|P}=[V_k+2nC^e_{Pk}-(2n+1)V_P]/(n+1)^2$. Provided $V_P>0$, $\Lambda^{EW}_{k|P}=(V_k+2nC^e_{Pk})/[(2n+1)V_P]<1$ if and only if $\Delta_{k|P}<0$.
\end{proposition}

$\Delta$ serves as the decision statistic; $\Lambda^{EW}$ is purely descriptive. Equal-weight admission addresses a particular question: does incorporating $A$ at equal weight reduce the risk of \emph{this} pool? Proposition~\ref{prop:scale} reveals what that question entails.

\begin{proposition}[Scale-free and scale-mismatch components]\label{prop:dsplit}
\begin{equation}\label{eq:dsplit}
\Delta_{A|P}=\underbrace{\rho_P^2-\rho_{P\cup A}^2}_{\Dsf_{A|P}}+\underbrace{q_{P\cup A}(1-c_{P\cup A})^2-q_P(1-c_P)^2}_{\Dsc_{A|P}},
\end{equation}
where, with $m=|A|$, pool mean alignments $\bar g_P,\bar g_A$, squared norms $q_P,q_A$ and $q_{PA}=w_P'\Cbar w_A$,
\begin{equation}\label{eq:dilution}
\rho^2_{P\cup A}=\frac{(n\bar g_P+m\bar g_A)^2}{n^2q_P+m^2q_A+2nm\,q_{PA}}.
\end{equation}
If $\bar g_P=0$ and $q_{PA}=0$, then $-\Dsf_{A|P}\le(m\bar g_A)^2/(n^2q_P)$.
\end{proposition}

This yields two implications. First, under weak alignment, $\Delta\approx\Dsc$: a candidate that shrinks the composite's norm, such as one that is negatively correlated with or orthogonal to the incumbent, reduces risk even when no alignment can be detected. Second, the scale-free benefit of an aligned batch decays like $(m/n)^2$, which means that equal-weight admission into a large pool is inherently unresponsive to aligned candidates. The \emph{scale-free rule} subjects $\Dsf$ to the three-way decision, with each composite retaining its own history scale. An envelope argument shows that estimating this scale influences the aggregated risk only to second order (\ref{app:proofs}). The scale-free rule determines whether a candidate lifts the pooled correlation of \emph{this equal-weight pool}. It is not a universal test of whether a candidate carries predictive value; from-scratch selection or regression-type weighting serve that purpose.

\section{Inference and evaluation design}\label{sec:selection}

\paragraph{Nested rolling origin} Following rolling-origin evaluation \citep{tashman2000}, each outer origin constructs its history from those dates whose targets are realized prior to the first test date. This history is split into two segments: a fit window and a trailing 52-date validation window. Tuning of the ridge penalties, the practical margins $\delta$ and $\delta^{\mathrm{sf}}$, the peLASSO penalty, and the shrinkage intensity proceeds by fitting on the fit window while scoring realized validation risk. Selection statistics are subsequently recomputed over the full history, after which the pools and weights are frozen for a 26-date outer test block. That block serves a single purpose: evaluation, together with the retrospective regret oracle.

\paragraph{Three-way rule} Consider an incumbent $P$ and a candidate family $\{A_j\}$. Constructing the date-level differences $d_{j,t}=\ell_{P\cup A_j,t}-\ell_{P,t}$ produces $\hat\Delta_j=\sum_ta_td_{j,t}$. The Newey--West standard errors employ lag $\lfloor4(T/100)^{2/9}\rfloor$ \citep{neweywest1987,neweywest1994}. Define the max-$t$ critical value $c$ as the $(1-\alpha)$ quantile of $\max_j|Z_j|$ evaluated under the estimated HAC correlation \citep{hothorn2008,romanowolf2005}. Truncation of its factor to eigenvalues exceeding a relative tolerance permits non-positive-definite estimates to be processed without repair. Admission occurs when $\hat\Delta_j+c\,\widehat{SE}_j<-\delta$; rejection occurs when $\hat\Delta_j-c\,\widehat{SE}_j>0$; otherwise the candidate remains undecided; $\alpha=0.05$. Deciding by the position of a simultaneous band relative to a practical margin follows the interval-inclusion logic of equivalence testing \citep{schuirmann1987}. The minimum detectable effect $\delta+(c+z_{0.8})\widehat{SE}_j$ is reported. Greedy selection commences from the forecaster exhibiting the highest history alignment (highest squared alignment under the scale-free rule), and each step admits the candidate whose $\hat\Delta$ is smallest. Termination occurs once no candidate is admitted.

\begin{remark}[What is and is not controlled]\label{rem:control}
Fix the incumbent $P$ and the candidate family. Provided the date-level differences satisfy a central limit theorem and the HAC estimator is consistent, the bands jointly cover all $\Delta_{A_j|P}$ with asymptotic probability $1-\alpha$. Consequently, the probability of admitting any candidate whose $\Delta_{A_j|P}\ge-\delta$ is asymptotically bounded by $\alpha/2$, \emph{conditionally on} $P$. Greedy selection, by contrast, renders the incumbent, the initial forecaster, and the family at subsequent steps all data-dependent. The per-step guarantee therefore fails to propagate to the path, and no selective-inference correction \citep{berk2013} is imposed. Assessment of path-level behavior is restricted to simulation (Section~\ref{sec:simulations}).
\end{remark}

\paragraph{Benchmarks}
\begin{itemize}
\item \emph{Selection:} greedy three-way admission using the equal-weight basis (with and without multiplicity adjustment) and the scale-free basis; greedy admission relying on point estimates; exhaustive search across all $2^N-1$ pools of history risk; size-matched selection via minimum mean deviation correlation, minimum mean forecast correlation, or maximum alignment; and the partially egalitarian LASSO \citep{dieboldshin2019}, that is, a nonnegative LASSO \citep{tibshirani1996} on the aggregated moments with an equal-weighted support.
\item \emph{Weighting:} equal weights; nonnegative, affine, and ridge-to-equal quadratic weights on shrunk moments; and the unconstrained ridge projection $(\Ctil+\eta I)^{-1}\gtil$.
\item \emph{Rescaling and null:} each pool and weight vector is additionally evaluated in a scale-calibrated version with its history scale held fixed, and against the no-information forecast.
\end{itemize}

\paragraph{Joint test of zero alignment} We assess $H_0\!:E[\gamma_{it}]=0$ for all $i$ using a HAC Wald statistic $\gbar'\widehat\Omega^{+}\gbar$, calibrated via a moving-block bootstrap \citep{kunsch1989} of the recentered statistic (block length 4, 999 draws), alongside a simultaneous max-$|t|$ test.

\paragraph{Criterion validity} For incumbents comprising the greedy path plus 30 random pools per origin, we calculate the within-incumbent Spearman correlation between each history criterion and the candidate's future incremental risk, as well as the pick regret associated with each criterion's top choice.

\paragraph{Evaluation} Our reported metrics include mean test risk with HAC intervals for differences from equal weighting and from the no-information forecast \citep{dieboldmariano1995,giacominiwhite2006}, the exact decomposition of Proposition~\ref{prop:risk}, composite correlation and rank information coefficients, a top-minus-bottom quintile return spread, weight diagnostics, and regret relative to the exhaustive test oracle.

The procedures and evaluation design were refined during the project using the same data. Consequently, tests across the 20 reported procedures are exposed to data snooping \citep{romanowolf2005} and are descriptive; we indicate Bonferroni thresholds where relevant.

\section{Simulations}\label{sec:simulations}

We generate forecasts according to $x_{it,m}=\sigma_i[\beta_{it}r_{t,m}+a_{0i}z_{t,m}+a_{1i}z^{c(i)}_{t,m}+\sqrt{\psi_i}\varepsilon_{it,m}]+\mu_i$, where $\beta_{it}=\beta_i+\phi\xi_t$. In this specification, $z$ represents a common factor, $z^c$ a cluster factor, $\xi_t$ a unit-variance AR(1) process, and $(\sigma_i,\mu_i)$ denote unequal scales that standardization eliminates. For each cell, the truth corresponds to the expected within-date geometry at identical $(N,M)$, estimated from 20{,}000 simulated dates; the resulting Monte Carlo standard error for mean correlations remains under 0.001. Each replication executes the complete nested procedure described in Section~\ref{sec:selection}, employing 156 history dates unless otherwise noted, alongside a 52-date validation window and a 52-date test block. Table~\ref{tab:simdesign} summarizes the designs, while parameter values and code are included in the replication package described under Data and code availability.

\begin{table}[t]\centering\small
\caption{Simulation designs ($N=24$, $M=60$ unless stated). Truth values: mean alignment $\bar\gamma$ and mean forecast correlation $\bar\rho^s$.}\label{tab:simdesign}
\resizebox{\textwidth}{!}{%
\begin{tabular}{@{}lp{12.2cm}@{}}
\toprule
Design & Content \\
\midrule
A & Recovery of $\Cbar,\gbar,\Kbar,\Kpool$ and PSD behavior under missing outputs; $N\in\{12,24\}$, $M\in\{20,30,60,240\}$, $T\in\{52,156,520\}$; 100 reps \\
B & Eight scenarios, 100 reps: S1 heterogeneous alignment with clusters ($\bar\gamma=0.091$, $\bar\rho^s=0.14$); S2 redundant near-clone clusters; S3 one third pure-noise forecasters; S4 exchangeable; S5 weak alignment ($\bar\gamma=0.022$) with heavy-tailed target; S6 time-varying alignment plus a break; S7 negatively aligned quarter; S8 S1 with 78 history dates \\
C & $N/M$ grid across the within-date rank boundary $M=N+2$; 50 reps \\
D & Phase diagram: alignment $\{0,0.03,0.08,0.15,0.25\}\times$ correlation $\{0,0.15,0.3,0.6\}\times N\in\{12,24\}$; 40 reps \\
E & History length $T\in\{52,104,260\}$ for S1, S2, S6; 60 reps \\
F & Instability mechanisms separated, 60 reps: F1 time-varying alignment only; F2 abrupt loss of the top quarter's alignment; F3 the same loss phased in over the test block; F4 dependence shift (common and cluster loadings rise, alignment loadings unchanged); F5 sign reversal of the top quarter's alignment \\
G & Misspecification of the generating geometry, 60 reps: G1 monotone nonlinear distortion of the scores before standardization; G2 cross-sectionally heteroskedastic, heavy-tailed target; G3 a test-period break hitting randomly chosen forecasters \\
\bottomrule
\end{tabular}}
\end{table}

\paragraph{Geometry recovery and dimensionality (Designs A, C)} Estimation errors decline approximately at the rate $1/\sqrt{TM}$. Joint shrinkage proves beneficial only when both $T$ and $M$ are small, reducing errors by 8--13\% at $(M,T)=(20,52)$. Employing $\Kpool$ rather than $\Kbar$ leaves a non-vanishing error as $T$ grows. Common-support matrices remained PSD across all replications. Pairwise deletion rendered 91--100\% of date-level matrices indefinite near the rank boundary, while shrinkage at the tuned intensity still left 5--26\% indefinite. Performance deteriorates smoothly through $M=N+2$ because aggregation restores full rank.

\paragraph{Joint tests} Under the null, the asymptotic HAC Wald test rejects in 20--38\% of replications when $N=12$ and in 53--70\% when $N=24$ (nominal 5\%). The bootstrap Wald test rejects in no more than 2.5\% of cases, and max-$|t|$ in 0--10\%. At alignment 0.03, their power ranges from 0.82 to 1.00.

\paragraph{Decisions and performance (Design B)} Table~\ref{tab:simB} presents decision rates and risk. Across scenarios, 2{,}200--6{,}600 non-improving candidate evaluations per procedure yielded at most 0.1\% false admissions. The one-sided 95\% Clopper--Pearson upper bound \citep{clopper1934} never surpasses 0.26\%. Recall, by contrast, differs markedly. The equal-weight rule leaves 10--31\% of truly improving candidates undecided; removing the multiplicity adjustment reduces this to 6--25\%, whereas the scale-free rule leaves 59--100\% undecided. After composites are scale-calibrated, the equal-weight rule, exhaustive search and ridge weights lie within 0.005 of one another outside the break scenario. The unconstrained ridge projection achieves the best performance or ties in every stationary scenario; under the break (S6), only the calibrated full equal-weight composite improves upon the no-information forecast.

\begin{table}[t]\centering\small
\caption{Design B: proportion of truly improving candidates remaining undecided, together with true test risk minus 1 (negative values indicate improvement over the no-information forecast). ``Scaled'': history-calibrated composite. Monte Carlo standard errors fall below 0.004.}\label{tab:simB}
\resizebox{\textwidth}{!}{%
\begin{tabular}{@{}lrrrrrrrr@{}}
\toprule
 & S1 & S2 & S3 & S4 & S5 & S6 & S7 & S8 \\
\midrule
\multicolumn{9}{@{}l}{\emph{Improving candidates left undecided}}\\
Three-way, equal weight & 0.23 & 0.10 & 0.29 & 0.24 & 0.27 & 0.24 & 0.22 & 0.31 \\
Three-way, unadjusted & 0.18 & 0.06 & 0.23 & 0.19 & 0.22 & 0.18 & 0.16 & 0.25 \\
Three-way, scale-free & 0.60 & 0.68 & 0.80 & 0.94 & 1.00 & 0.70 & 0.59 & 0.77 \\
\multicolumn{9}{@{}l}{\emph{True test risk minus 1}}\\
Equal weight, all & $-$0.004 & 0.074 & 0.010 & 0.129 & 0.230 & 0.072 & 0.049 & $-$0.006 \\
Three-way, equal weight & $-$0.022 & 0.008 & 0.030 & 0.173 & 0.248 & 0.252 & $-$0.023 & $-$0.013 \\
Equal weight, all (scaled) & $-$0.046 & $-$0.021 & $-$0.020 & $-$0.024 & $-$0.002 & $-$0.008 & $-$0.022 & $-$0.047 \\
Three-way, equal weight (scaled) & $-$0.079 & $-$0.053 & $-$0.026 & $-$0.021 & $-$0.007 & 0.069 & $-$0.080 & $-$0.078 \\
Three-way, scale-free (scaled) & $-$0.070 & $-$0.042 & $-$0.022 & $-$0.009 & $-$0.003 & 0.063 & $-$0.071 & $-$0.058 \\
Ridge weights (scaled) & $-$0.081 & $-$0.053 & $-$0.028 & $-$0.024 & $-$0.005 & 0.025 & $-$0.081 & $-$0.081 \\
Ridge projection & $-$0.101 & $-$0.215 & $-$0.038 & $-$0.024 & $-$0.019 & 0.050 & $-$0.224 & $-$0.100 \\
\bottomrule
\end{tabular}}
\end{table}

\begin{figure}[t]\centering
\includegraphics[width=\textwidth]{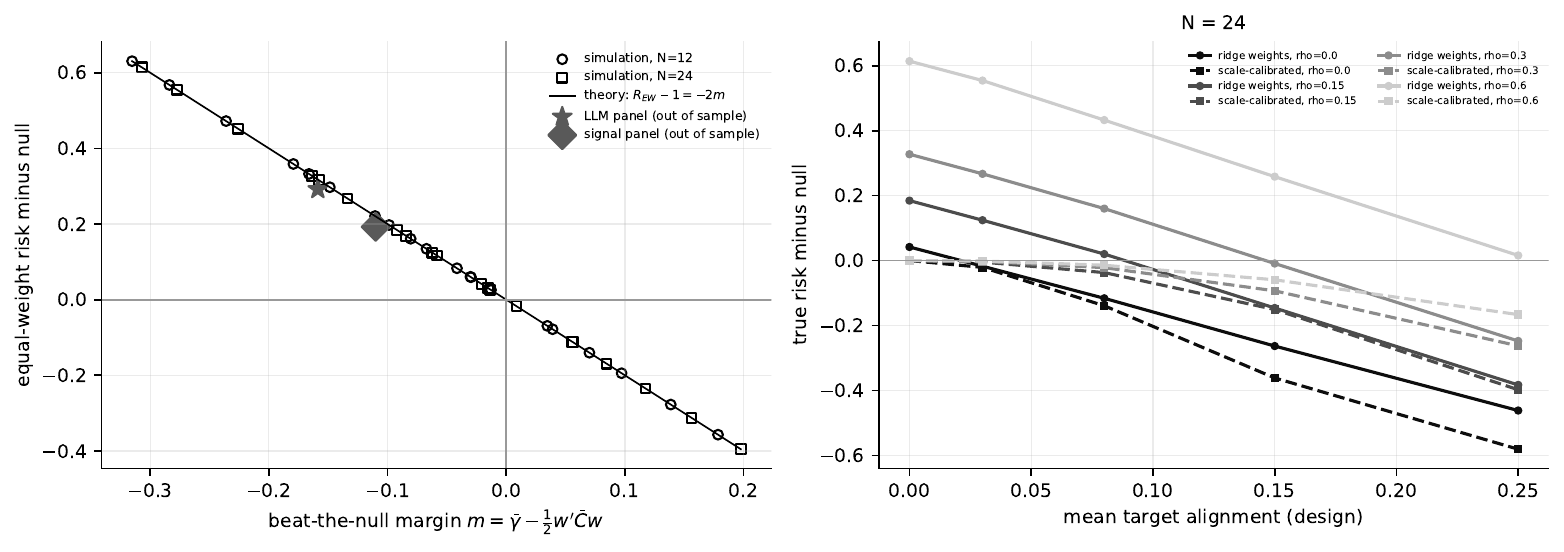}
\caption{Left: equal-weight true risk minus 1 plotted against the margin $m$ of Corollary~\ref{cor:null} for all Design~D cells, displaying the exact line $-2m$ and the two empirical panels (out of sample). Right: ridge weights alongside their scale-calibrated version across alignment levels ($N=24$).}\label{fig:phase}
\end{figure}

\paragraph{When can combination help? (Design D)} Every Design~D cell is located by Figure~\ref{fig:phase} along the exact line $\Rrel(\one/N)-1=-2m$ derived in Corollary~\ref{cor:null}. Only in regions where alignment stands sufficiently high relative to dependence does equal weighting outperform the no-information forecast. With alignment at 0.03 and zero correlation, the strategy sacrifices 0.026 when $N=12$ yet gains 0.017 when $N=24$—precisely what the $(1-\bar\rho)/N$ term predicts. When alignment reaches 0.15 alongside correlation 0.6 ($N=24$), the attainable improvement equals $-0.136$; the ridge projection nearly achieves this at $-0.132$, while calibrated selections together with ridge weights attain approximately $-0.06$.

\begin{figure}[t]\centering
\includegraphics[width=\textwidth]{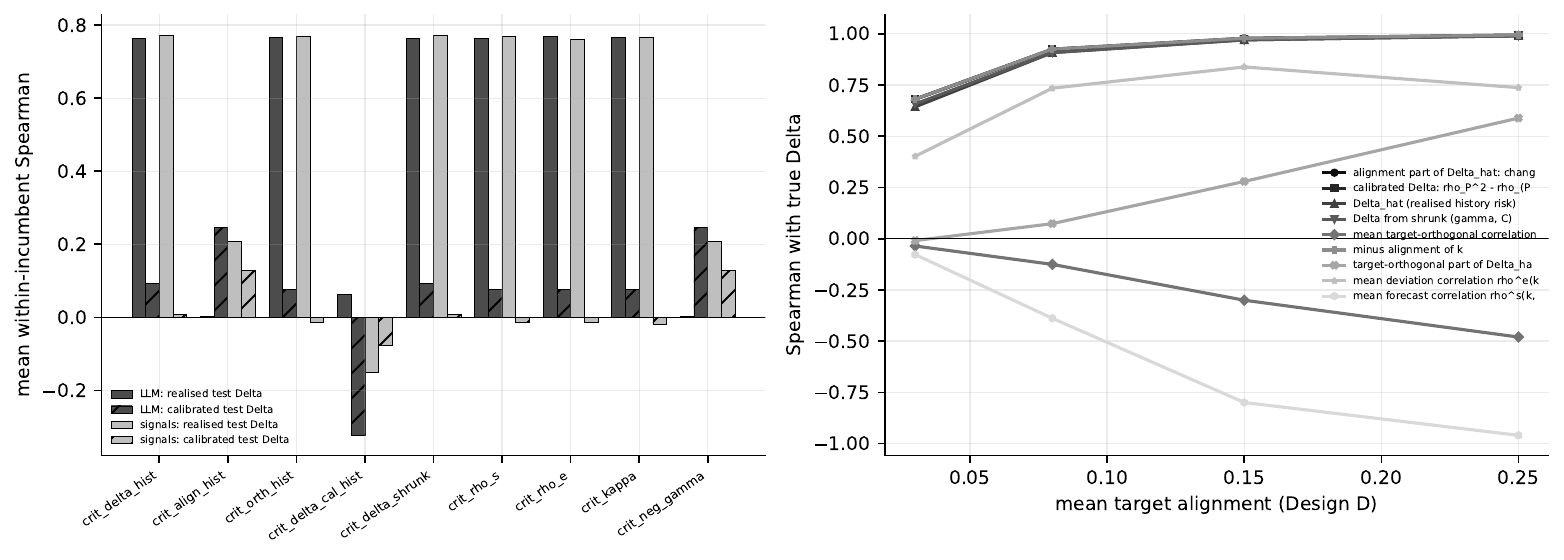}
\caption{Criterion validity. Left: LLM and signal panels versus realized and calibrated test outcomes. Right: Design~D versus true incremental risk.}\label{fig:crit}
\end{figure}

\paragraph{Which history statistic ranks candidates? (Designs B, D)} Criterion validity results appear in Table~\ref{tab:crit} and Figure~\ref{fig:crit}.
\begin{itemize}
\item \textbf{Exchangeable dependence (Design D).} Once alignment reaches at least 0.08, $\hat\Delta$ achieves near-perfect ranking of candidates according to their true incremental risk. This ranking derives from its alignment component; alignment alone performs equally well.
\item \textbf{Forecast correlation ranks backwards.} Its Spearman correlation ranges from $-0.39$ to $-0.96$. The identity $C=\gamma\gamma'+K$ implies that better-aligned forecasters are mechanically more correlated with each other, rendering the ``most diverse'' candidate the least aligned. The target-orthogonal correlation likewise ranks backwards, since its normalization $K_{ii}=1-\gamma_i^2$ incorporates alignment.
\item \textbf{Heterogeneous dependence (Design B).} $\hat\Delta$ dominates: 0.98 in S2 versus 0.51 for alignment alone, and 0.87 in S3 versus 0.20. Outside the break scenario, its pick regret remains at most 0.002, compared with up to 0.074 for forecast correlation.
\end{itemize}

\begin{table}[t]\centering\small
\caption{Criterion validity: mean within-incumbent Spearman correlation between history criteria and the true incremental risk (higher values indicate better performance). Design~D results are averaged across dependence levels and $N$.}\label{tab:crit}
\resizebox{\textwidth}{!}{%
\begin{tabular}{@{}lrrrrrrrr@{}}
\toprule
 & \multicolumn{4}{c}{Design D: alignment} & \multicolumn{4}{c}{Design B: scenario} \\
\cmidrule(lr){2-5}\cmidrule(l){6-9}
Criterion & 0.03 & 0.08 & 0.15 & 0.25 & S1 & S2 & S3 & S5 \\
\midrule
$\hat\Delta$ (realized history risk) & 0.64 & 0.91 & 0.97 & 0.99 & 0.97 & 0.98 & 0.87 & 0.94 \\
\quad alignment part & 0.68 & 0.92 & 0.98 & 0.99 & 0.89 & 0.51 & 0.20 & 0.69 \\
\quad target-orthogonal part & $-$0.01 & 0.07 & 0.28 & 0.59 & 0.41 & 0.73 & 0.39 & 0.51 \\
Scale-free $\hat\Delta^{\mathrm{sf}}$ & 0.68 & 0.92 & 0.97 & 0.99 & 0.93 & 0.82 & 0.37 & 0.70 \\
Mean deviation correlation $\rho^e(k,P)$ & 0.40 & 0.73 & 0.84 & 0.74 & 0.87 & 0.96 & 0.65 & 0.84 \\
Mean forecast correlation $\rho^s(k,P)$ & $-$0.08 & $-$0.39 & $-$0.80 & $-$0.96 & 0.00 & 0.50 & 0.30 & 0.46 \\
Mean orthogonal correlation & $-$0.04 & $-$0.13 & $-$0.30 & $-$0.48 & 0.28 & 0.65 & 0.36 & 0.50 \\
Alignment of $k$ & 0.68 & 0.92 & 0.98 & 0.99 & 0.89 & 0.51 & 0.20 & 0.69 \\
\bottomrule
\end{tabular}}
\end{table}

\paragraph{Misspecification (Design G)} Although the identities hold irrespective of the generating process, the procedures could nevertheless be tuned to it. Under a monotone nonlinear distortion of the scores and under a heteroskedastic, heavy-tailed target, the ordering remains unchanged: the ridge projection achieves the best performance ($-0.095$ and $-0.131$ relative to the no-information forecast, compared with attainable values of $-0.098$ and $-0.133$), calibrated selections attain $-0.066$ to $-0.079$, and calibrated equal weighting reaches $-0.046$ to $-0.061$. When a break strikes randomly chosen forecasters rather than the most aligned ones, every calibrated method remains slightly below the no-information forecast ($-0.013$ to $-0.026$), whereas the unscaled equal-weight pool lies above it ($+0.059$). Criterion validity is likewise unaffected: $\hat\Delta$ achieves the best candidate ranking (Spearman 0.97, 0.98 and 0.40), while forecast correlation proves uninformative or negative ($0.02$, $-0.09$, $0.16$).

\paragraph{History length and instability (Designs E, F)} Table~\ref{tab:simEF} presents the following findings.
\begin{itemize}
\item \textbf{Stationary designs.} At every history length, even 52 dates, the ridge projection exhibits the lowest risk. The calibrated equal-weight rule, peLASSO, exhaustive search and ridge weights lie within approximately 0.005 of one another, and the scale-free rule improves as $T$ grows.
\item \textbf{Projection weights.} In S2, the projection's advantage relies on large offsetting weights across near-clones: gross exposure 5.1--5.4, with half of the weights negative and maximum absolute weight 0.39. Its gross exposure in S1 equals 1.04--1.16.
\item \textbf{Instability mechanisms matter.}
\begin{itemize}
\item Time-varying alignment without a break (F1) behaves similarly to the stationary case.
\item When the top forecasters abruptly lose alignment (F2) or reverse it (F5), every estimated weighting and selection performs worse than the no-information forecast. The calibrated full equal-weight composite proves least harmful (F2 constitutes the only case in which it remains below it).
\item When the loss occurs gradually (F3), calibrated equal weights, ridge weights, exhaustive search and the projection perform comparably.
\item When only dependence shifts (F4), the projection achieves the best result ($-0.051$), while the calibrated full equal-weight composite performs slightly worse than the no-information forecast ($+0.003$).
\end{itemize}
\end{itemize}
No single rule proves robust to all mechanisms. The geometry identifies which term moved: alignment breaks harm estimated weights, whereas dependence shifts harm equal weighting. The former pattern agrees with the case for pooling under structural breaks \citep{hendryclements2004}.

\begin{table}[t]\centering\small
\caption{Designs E and F: true test risk minus 1. All pools and weights are scale-calibrated except the unconstrained ridge projection. ``Attainable'': Corollary~\ref{cor:attain} under the test-period truth. 60 replications; Monte Carlo standard errors fall below 0.004.}\label{tab:simEF}
\resizebox{\textwidth}{!}{%
\begin{tabular}{@{}lrrrrrrrrr@{}}
\toprule
 & \multicolumn{2}{c}{E: S1} & \multicolumn{2}{c}{E: S2} & \multicolumn{5}{c}{F: instability mechanism} \\
\cmidrule(lr){2-3}\cmidrule(lr){4-5}\cmidrule(l){6-10}
 & $T=52$ & $T=260$ & $T=52$ & $T=260$ & F1 & F2 & F3 & F4 & F5 \\
\midrule
Equal weight, all & $-$0.046 & $-$0.046 & $-$0.021 & $-$0.021 & $-$0.043 & $-$0.010 & $-$0.028 & 0.003 & 0.028 \\
Three-way, equal weight & $-$0.074 & $-$0.080 & $-$0.049 & $-$0.053 & $-$0.077 & 0.068 & $-$0.006 & $-$0.022 & 0.222 \\
Three-way, scale-free & $-$0.047 & $-$0.076 & $-$0.019 & $-$0.048 & $-$0.061 & 0.068 & $-$0.001 & $-$0.023 & 0.212 \\
peLASSO & $-$0.077 & $-$0.080 & $-$0.052 & $-$0.053 & $-$0.076 & 0.034 & $-$0.023 & $-$0.013 & 0.154 \\
Exhaustive search & $-$0.077 & $-$0.078 & $-$0.053 & $-$0.053 & $-$0.075 & 0.010 & $-$0.034 & $-$0.009 & 0.103 \\
Ridge weights & $-$0.078 & $-$0.082 & $-$0.052 & $-$0.053 & $-$0.078 & 0.025 & $-$0.029 & $-$0.007 & 0.134 \\
Ridge projection & $-$0.096 & $-$0.102 & $-$0.211 & $-$0.216 & $-$0.098 & 0.050 & $-$0.026 & $-$0.051 & 0.210 \\
\midrule
Attainable (truth) & & & & & $-$0.103 & $-$0.073 & $-$0.056 & $-$0.076 & $-$0.256 \\
\bottomrule
\end{tabular}}
\end{table}

\section{A retrospective stress test with language-model forecasters}\label{sec:empirical}

\subsection{Design and its limits}
The ensemble crosses four base models drawn from four developer lineages (\texttt{gpt-5-nano}, \texttt{deepseek-v4-flash}, \texttt{Llama-3.1-8B-Instruct}, \texttt{gemini-3.5-flash-lite}) with three personas (momentum, value reversal, macro defensive) and two information subsets. The price-only subset comprises trailing 1-, 3- and 12-month returns; the price-plus-volatility subset additionally includes 63-day volatility and 6-month maximum drawdown. This yields $N=24$ forecasters. At each weekly date, every forecaster receives a point-in-time cross-section of 60 US large-capitalization equities and assigns scores in $[-1,1]$ to every ticker as a relative ranking of the forward five-day return.

Three features of the design constrain what it can demonstrate.
\begin{itemize}
\item \textbf{Retrospective forecasts.} All forecasts were produced retrospectively in 2026 from point-in-time inputs. Because they constitute outputs of models whose training data may postdate the forecast dates, they do not represent real-time forecasts \citep[on look-ahead bias in LLM-based return prediction, see][]{glasserman2024}. The stated training cutoffs are December 2023 (\texttt{Llama}) and May 2024 (\texttt{gpt}); the remaining two models state March 2026 (\texttt{gemini}) and April 2026 (\texttt{deepseek}), corresponding to the end of the sample.
\item \textbf{Fixed universe.} The equity universe was fixed once, at the end of the sample, so the cross-section is conditioned on survival \citep{brown1992}. The application constitutes a fixed-panel ranking experiment, not an investable backtest.
\item \textbf{Scope.} Prompts, decoding settings, response hashes and access timestamps are included in the replication package described under Data and code availability.
\end{itemize}
We employ the panel as a stress test of the framework: a case featuring many forecasters, strong redundancy and weak alignment, in which the geometry, admission rules and power can be examined. It does not constitute evidence about real-time LLM forecasting ability; in other settings, aggregated LLM forecasts have rivaled human crowds \citep{schoenegger2024}.

Of 285 weekly dates from January 2021 to June 2026, 261 enter the fixed-membership panel. The 24 exclusions comprise dates on which at least one of the 6{,}840 forecaster-date cells failed (21 constant outputs, 4 unparsable responses). The failures concentrate in one model (19 of 25 failed cells), one persona (20) and the price-plus-volatility subset (17). They show no relation to market conditions: excluded and retained dates exhibit similar cross-sectional return dispersion (0.038 against 0.036, $p=0.65$), mean return ($p=0.82$), absolute mean return ($p=0.40$) and trailing market volatility (0.160 against 0.152, $p=0.61$). The common support ranges from 49 to 60 tickers (median 60). Six nested outer origins with 104 initial history dates, 52 validation dates and 26-date test blocks yield 157 test dates (March 2023 to June 2026); the residual date at the end of the sample is absorbed into the final block, which therefore covers 27.

\subsection{Geometry}
\begin{figure}[t]\centering
\includegraphics[width=0.92\textwidth]{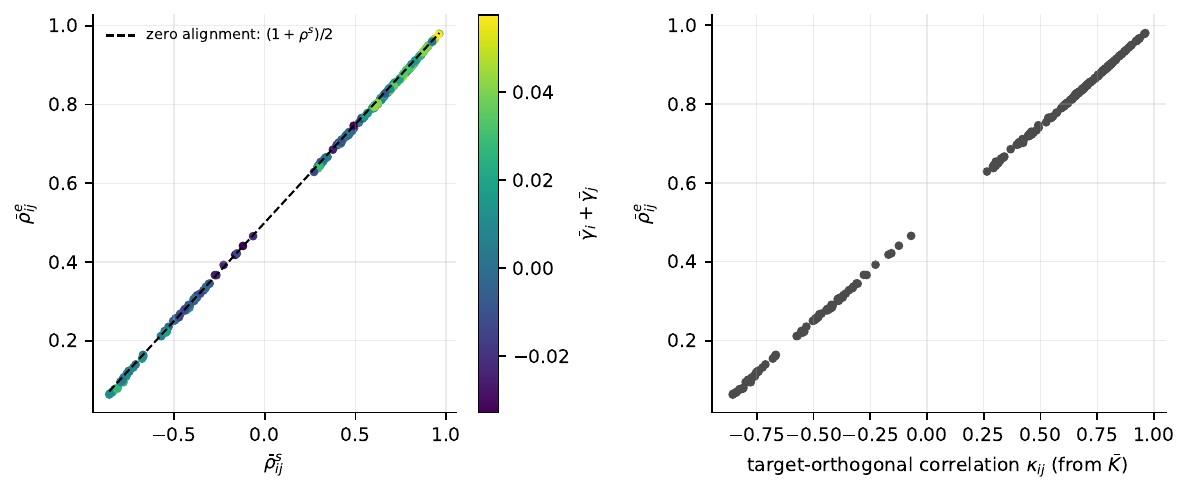}
\caption{Pairwise deviation correlation plotted against forecast correlation (left, with the zero-alignment line) and against target-orthogonal correlation (right).}\label{fig:pairs}
\end{figure}

Table~\ref{tab:geometry} presents the geometry together with 95\% moving-block bootstrap intervals.
\begin{itemize}
\item \textbf{Alignment is negligible.} Mean alignment equals $0.006$ $[-0.010,0.020]$.
\item \textbf{Deviation correlation is a translation of forecast correlation.} Mean forecast correlation equals $0.300$ and mean deviation correlation $0.645$, lying within $0.005$ of the zero-alignment benchmark $0.650$. Across the 276 pairs, deviation correlation constitutes almost exactly a linear function of forecast correlation ($R^2=0.9999$; Figure~\ref{fig:pairs}): 29\% of forecast correlations are negative, yet no deviation correlation is. The equal-weight variance-equivalent size equals 3.03 from forecast correlation and 1.52 from deviation correlation.
\item \textbf{$\Kbar$ and $\Kpool$ differ.} Time variation in alignment accounts for 5.9\% of the trace of $\Kpool$, so the two decompositions of equal-weight risk differ ($1.010+0.308$ within dates, $0.988+0.330$ pooled).
\item \textbf{Dependence is concentrated.} The leading eigenvalue of $\Cbar$ carries 66\% of its trace. Descriptively, it loads on the momentum and macro-defensive personas, and same-persona pairs correlate at $0.47$ versus $0.23$ otherwise.
\item \textbf{No evidence of alignment anywhere in the ensemble.} The bootstrap Wald test yields $p=0.065$ and max-$|t|$ yields $p=0.25$; the asymptotic Wald $p$-value of 0.0001 is invalid (Section~\ref{sec:simulations}). Equal weighting misses the condition of Corollary~\ref{cor:null} by $m=-0.159$ $[-0.177,-0.142]$, and the in-sample attainable risk of any linear combination equals 0.994.
\end{itemize}

\begin{table}[t]\centering\small
\caption{Geometry of the LLM panel (261 dates, $N=24$); 95\% moving-block bootstrap intervals (block length 4, 999 draws).}\label{tab:geometry}
\resizebox{\textwidth}{!}{%
\begin{tabular}{@{}lrl@{}}
\toprule
Quantity & Estimate & 95\% interval \\
\midrule
Mean forecast correlation $\bar\rho^s$ / deviation correlation $\bar\rho^e$ & 0.300 / 0.645 & [0.284, 0.317] / [0.637, 0.654] \\
$\bar\rho^e-(1+\bar\rho^s)/2$ & $-$0.005 & [$-$0.008, $-$0.002] \\
Mean target alignment $\gbar$ & 0.006 & [$-$0.010, 0.020] \\
$N_{\mathrm{eff}}$ from $\bar\rho^s$ / from $\bar\rho^e$ & 3.03 / 1.52 & [2.89, 3.19] / [1.50, 1.53] \\
Trace share of $\operatorname{Cov}_a(\gamma_t)$ in $\Kpool$ & 0.059 & [0.049, 0.065] \\
Equal-weight risk / margin $m$ & 1.318 / $-$0.159 & [1.285, 1.354] / [$-$0.177, $-$0.142] \\
In-sample attainable risk; joint zero-alignment $p$ (bootstrap Wald / max-$|t|$) & 0.994; 0.065 / 0.25 & \\
\bottomrule
\end{tabular}}
\end{table}

\subsection{Selection and out-of-sample risk}
\paragraph{Selected pools}
\begin{itemize}
\item At every origin, the equal-weight three-way rule selects exactly two forecasters. The pair always exhibits negative correlation (mean within-pool $\rho^s=-0.78$) and remains identical under every multiplicity and $\alpha$ variant.
\item The rule's decisions prove independent of the practical margin. For every $\delta$ from $0.0001$ to $0.1$, the same pool is selected, and the scale-free rule retains only its starting forecaster for every $\delta^{\mathrm{sf}}$ from $10^{-5}$ to $0.01$. The margin is therefore not identified in this application, and no result depends on it.
\item Exhaustive search selects six forecasters, while peLASSO selects 3.3 on average.
\item Although selected pools are unstable (bootstrap Jaccard similarity 0.35--0.49), the risk surface is flat: at each origin, 19--57 pools lie within 0.005 of the history optimum, and that optimum (1.037--1.052) exceeds the no-information forecast at every origin.
\end{itemize}

\paragraph{Out-of-sample risk} Table~\ref{tab:oos} presents the following findings.
\begin{itemize}
\item \textbf{Against equal weighting,} quadratic weightings and the pools chosen by exhaustive search, three-way admission and diversity criteria reduce risk by 0.20--0.27.
\item \textbf{Against the no-information forecast,} none improves. The best performer, shrunk ridge weights, exhibits a difference of $+0.019$ $[0.006,0.032]$.
\item \textbf{The gains come from the orthogonal term.} The target-orthogonal term falls from 0.301 to 0.031--0.105, while the alignment term remains between 0.975 and 0.991.
\item \textbf{Scale-free diagnostics agree.} Composite correlations range from 0.025 to 0.043, and no method differs from equal weighting ($|t|\le0.84$). Top-minus-bottom quintile return spreads range from $-0.10\%$ to $0.50\%$ per five days, with no $|t|$ exceeding the Bonferroni threshold of 3.0 for 20 procedures. History-calibrated composites sit at the no-information forecast (0.9981--1.0014).
\item \textbf{The regression-type benchmarks effectively abstain.} The ridge projection's weights sum to 0.018 with gross exposure 0.11, and the raw affine weights carry 18\% negative mass.
\end{itemize}

\begin{table}[t]\centering\small
\caption{Out-of-sample results (157 test dates, six origins). Differences in relative-score risk with 95\% HAC intervals; ``Align.'' and ``Orth.'' decompose risk exactly; spread: top-minus-bottom quintile forward return of the composite (\% per five days, HAC $t$).}\label{tab:oos}
\resizebox{\textwidth}{!}{%
\begin{tabular}{@{}lrrllrrr@{}}
\toprule
Method & Size & Risk & vs equal weight & vs no-information & Align. & Orth. & Spread ($t$) \\
\midrule
Equal weight, all & 24 & 1.292 & -- & 0.292 [0.247, 0.337] & 0.991 & 0.301 & 0.36 (1.4) \\
Three-way, equal weight & 2.0 & 1.079 & $-$0.213 [$-$0.262, $-$0.164] & 0.079 [0.055, 0.103] & 0.978 & 0.101 & 0.19 (1.0) \\
Pairwise batch & 2.3 & 1.063 & $-$0.229 [$-$0.276, $-$0.182] & 0.063 [0.042, 0.084] & 0.975 & 0.088 & 0.23 (1.1) \\
peLASSO & 3.3 & 1.247 & $-$0.045 [$-$0.160, 0.069] & 0.247 [0.123, 0.371] & 0.986 & 0.260 & 0.50 (2.1) \\
Exhaustive search & 6.0 & 1.033 & $-$0.259 [$-$0.298, $-$0.220] & 0.033 [0.016, 0.050] & 0.986 & 0.047 & 0.24 (1.1) \\
Min.\ deviation correlation & 2.0 & 1.089 & $-$0.203 [$-$0.249, $-$0.157] & 0.089 [0.062, 0.116] & 0.984 & 0.105 & 0.17 (0.8) \\
Ridge weights (shrunk) & -- & 1.019 & $-$0.273 [$-$0.315, $-$0.230] & 0.019 [0.006, 0.032] & 0.988 & 0.031 & 0.14 (0.6) \\
Affine weights (raw) & -- & 1.020 & $-$0.272 [$-$0.316, $-$0.229] & 0.020 [0.006, 0.034] & 0.985 & 0.035 & 0.24 (1.0) \\
Ridge projection & -- & 1.000 & $-$0.292 & 0.000 [$-$0.001, 0.002] & 1.000 & 0.001 & $-$0.10 ($-$0.5) \\
No-information forecast & 0 & 1.000 & $-$0.292 [$-$0.337, $-$0.247] & -- & 1.000 & 0.000 & -- \\
\bottomrule
\end{tabular}}
\end{table}

\begin{figure}[t]\centering
\includegraphics[width=\textwidth]{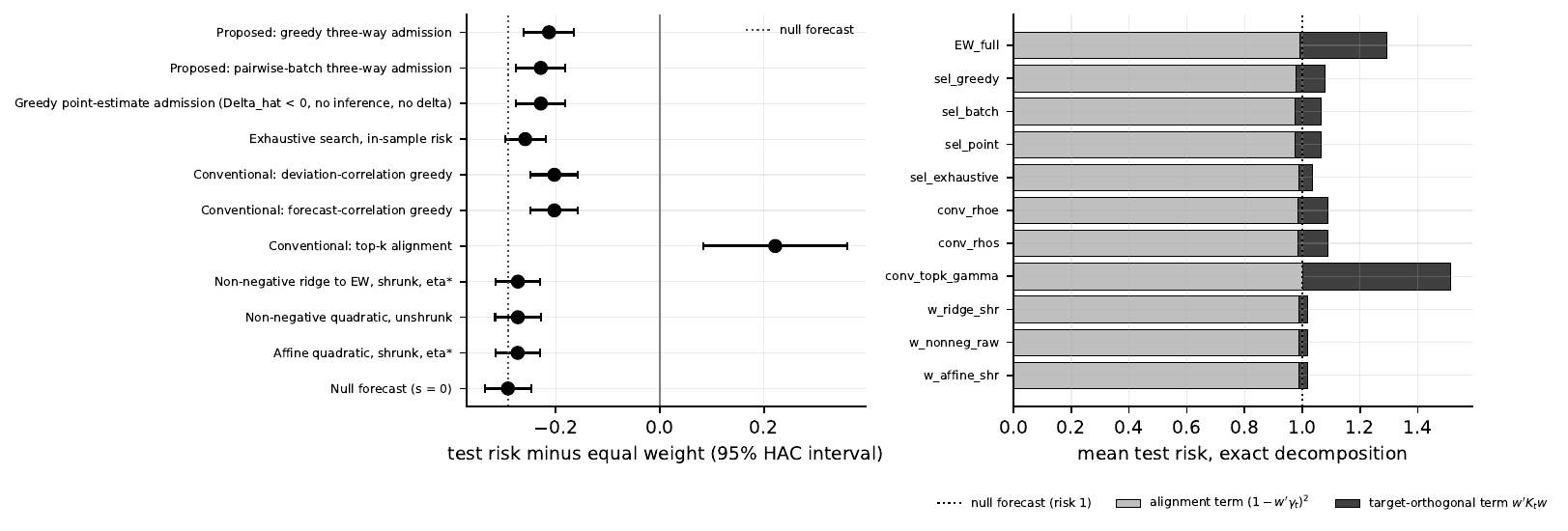}
\caption{Left: test risk minus equal-weight risk with 95\% HAC intervals (dotted: no-information forecast). Right: exact decomposition into alignment and target-orthogonal terms.}\label{fig:oos}
\end{figure}

\subsection{Control panel and cross-panel admission}
To distinguish uninformative forecasters from an uninformative feature set, we construct nine mechanical signals using exactly the features provided to the LLMs: 1-month reversal \citep{jegadeesh1990}, 3-, 12- and 12--1-month momentum \citep{jegadeeshtitman1993}, low volatility \citep{ang2006}, low drawdown, and three persona-emulation rules. These signals likewise exhibit no detectable alignment (mean 0.003, largest $|t|$ 1.82, bootstrap Wald $p=0.22$, attainable risk 0.997), and no combination of them improves upon the no-information forecast. We thus find no detectable alignment for this feature set and these deterministic transformations in this panel. This does not demonstrate that the target is unforecastable from other information, transformations or conditioning variables.

The comparison also characterizes the forecasters descriptively. Momentum-persona forecasts correlate $0.74$ with the momentum-emulation rule and $-0.82$ with the reversal rule. Macro-defensive forecasts behave similarly to momentum ($0.61$ and $-0.73$). Value-reversal forecasts are essentially uncorrelated with every rule (mean $0.05$).

The admission experiment adds LLM batches (by lineage, persona and information subset, and all 24) to the equal-weight signal pool, and signal batches to the LLM pool, at each origin (Table~\ref{tab:admission}).
\begin{itemize}
\item \textbf{Equal-weight basis.} At all six origins, the value-reversal batch is admitted into the signal pool with $\hat\Delta=-0.130$, of which the scale-free component equals $-0.0001$ and the scale mismatch $-0.130$. Its test $\Delta$ equals $-0.114$ ($t=-9.8$): pure dilution by forecasts uncorrelated with the incumbent.
\item \textbf{Scale-free basis.} Every batch in both directions remains undecided at every origin, and realized scale-free test $\Delta$ lies within $\pm0.0005$.
\end{itemize}

\begin{table}[t]\centering\small
\caption{Cross-panel admission (six origins; simultaneous max-$t$ bands over each family; $\delta=0.005$, $\delta^{\mathrm{sf}}=0.0005$). Test $\Delta$ pooled over 157 test dates (HAC $t$).}\label{tab:admission}
\resizebox{\textwidth}{!}{%
\begin{tabular}{@{}llrrrrr@{}}
\toprule
Candidate $\to$ incumbent & Basis & $\hat\Delta$ & of which $\Dsf$ & Admit & Reject & Test $\Delta$ ($t$) \\
\midrule
Value-reversal LLMs (8) $\to$ signals & equal weight & $-$0.130 & $-$0.000 & 6/6 & 0/6 & $-$0.114 ($-$9.8) \\
Momentum LLMs (8) $\to$ signals & equal weight & 0.072 & $-$0.000 & 0/6 & 4/6 & 0.087 (4.7) \\
All 24 LLMs $\to$ signals & equal weight & $-$0.001 & $-$0.000 & 0/6 & 0/6 & 0.008 (0.4) \\
All nine signals $\to$ LLMs & equal weight & $-$0.097 & $-$0.000 & 6/6 & 0/6 & $-$0.090 ($-$11.5) \\
Any batch, either direction & scale-free & [$-$0.0001, 0.0000] & -- & 0/6 & 0/6 & [$-$0.0003, 0.0005] \\
\bottomrule
\end{tabular}}
\end{table}

\subsection{Positive control: what could have been detected}
\begin{figure}[t]\centering
\includegraphics[width=\textwidth]{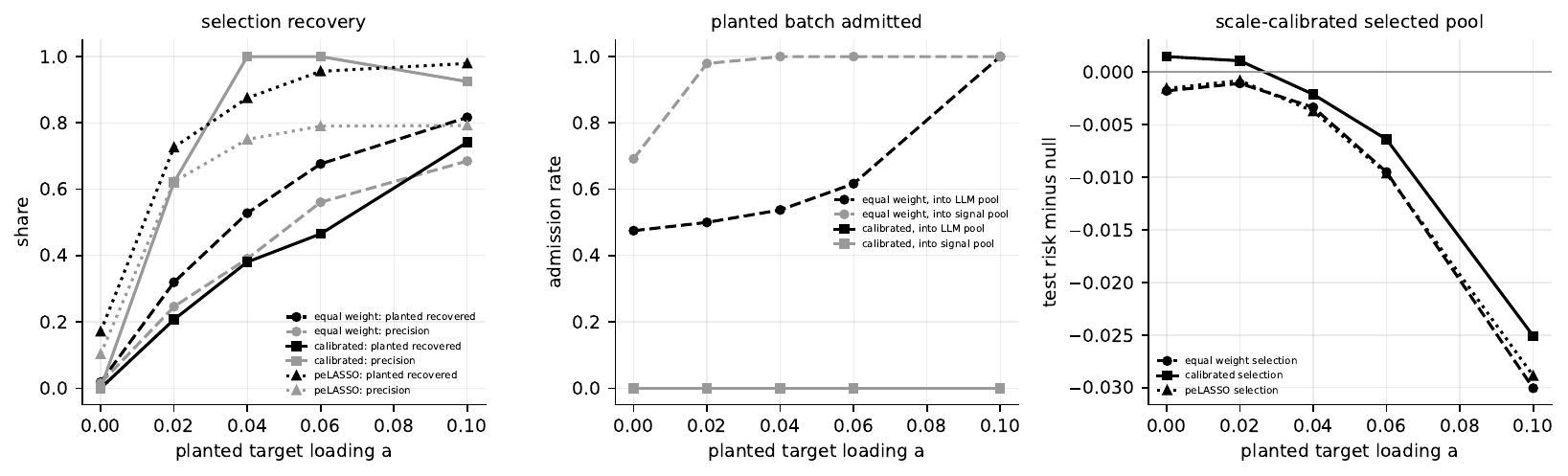}
\caption{Positive control: recovery and precision (left panel), admission of the planted batch into large pools (middle panel), and risk of calibrated selections (right panel).}\label{fig:positive}
\end{figure}

Only if signal could have been detected does a null result carry information. Three forecasters are planted per replicate, $p_{jt}=a\,y_t+0.7\,u^\perp_{jt}+\sqrt{0.51-a^2}\,e_{jt}$. In this specification, $u^\perp$ denotes the target-orthogonal component of a randomly drawn real LLM forecast, thereby generating real dependence and exactly zero alignment at $a=0$, while $e$ represents noise (40 replicates, six origins). Because the planted forecasters employ the realized target, they serve solely as a validation device. Table~\ref{tab:positive} and Figure~\ref{fig:positive} present the results.
\begin{itemize}
\item \textbf{Admission into large pools.} Discrimination proves impossible for equal-weight admission: zero-loading planted forecasters are admitted 24\% of the time into the LLM pool and 62\% into the signal pool. Even at $a=0.10$, scale-free admission never admits into either large pool, consistent with the dilution bound (three candidates with alignment 0.10 added to 24 incumbents gain at most about $0.0005$ in squared correlation).
\item \textbf{Selection from scratch} exhibits a precision--recall trade-off.
\begin{itemize}
\item Essentially nothing is recovered by the scale-free rule at $a=0$. At $a=0.04$, $0.06$ and $0.10$, recovery rates reach 38\%, 49\% and 77\% of planted forecasters with precision 1.00, 1.00 and 0.94.
\item Recovery rates for the equal-weight rule range from 48--82\% with precision 0.36--0.69.
\item peLASSO achieves recovery rates of 89--99\% with precision 0.76--0.82, but includes 12\% of zero-loading forecasters.
\end{itemize}
\item \textbf{Risk.} Similar performance is exhibited by the calibrated composites of the three selections: 0.027--0.031 below the no-information forecast at $a=0.10$, and at it when $a=0$.
\item \textbf{Alignment that varies over time is harder to find.} With autocorrelated alignment of the same mean, recovery remains essentially unchanged (0.42 at $a=0.06$ for the scale-free rule). When the planted alignment is present only in the second half of the sample, recovery falls to 0.19 at $a=0.06$ and 0.25 at $a=0.10$, with precision 0.57--0.66. Detection power therefore constitutes a statement about stable alignment.
\end{itemize}

\begin{table}[t]\centering\small
\caption{Positive control: selection from scratch in the LLM-plus-planted universe (40 replicates $\times$ 6 origins). Recovery denotes the share of planted forecasters selected; precision denotes the planted share of the selected pool; risk denotes the calibrated selected pool minus the no-information forecast.}\label{tab:positive}
\resizebox{\textwidth}{!}{%
\begin{tabular}{@{}lrrrrrrrrr@{}}
\toprule
 & \multicolumn{3}{c}{Three-way, scale-free} & \multicolumn{3}{c}{Three-way, equal weight} & \multicolumn{3}{c}{peLASSO} \\
\cmidrule(lr){2-4}\cmidrule(lr){5-7}\cmidrule(l){8-10}
Loading $a$ & Recovery & Precision & Risk & Recovery & Precision & Risk & Recovery & Precision & Risk \\
\midrule
0.00 & 0.00 & 0.00 & 0.002 & 0.02 & 0.01 & $-$0.002 & 0.12 & 0.08 & $-$0.002 \\
0.02 & 0.21 & 0.62 & 0.001 & 0.32 & 0.25 & $-$0.001 & 0.74 & 0.62 & $-$0.001 \\
0.04 & 0.38 & 1.00 & $-$0.002 & 0.48 & 0.36 & $-$0.003 & 0.89 & 0.76 & $-$0.004 \\
0.06 & 0.49 & 1.00 & $-$0.007 & 0.62 & 0.52 & $-$0.010 & 0.96 & 0.81 & $-$0.009 \\
0.10 & 0.77 & 0.94 & $-$0.027 & 0.82 & 0.69 & $-$0.031 & 0.99 & 0.82 & $-$0.031 \\
\bottomrule
\end{tabular}}
\end{table}

\subsection{Placebo, per-origin results and effective sample size}
What the null can support is constrained by three diagnostics.
\begin{itemize}
\item \textbf{Placebo.} Forecast dependence is preserved exactly, and alignment destroyed, by circularly shifting the realized target through random offsets (50 shifts). Mean forecast correlation remains unchanged at $0.3005$, while the alignment statistics of the actual data sit in the middle of the placebo distribution: mean placebo alignment equals $0.0042$ versus $0.0061$ actual, and mean placebo $\max|t|$ equals $2.29$ (range $0.77$ to $5.10$) versus $2.13$ actual. No distributional assumption is required for the resulting placebo-calibrated $p$-values, which equal $0.55$ for mean alignment, $0.53$ for $\max|t|$, $0.22$ for the Wald statistic and $0.61$ for the equal-weight composite $t$. At the nominal 5\% level, rejection occurs in 22\% of placebos for the bootstrap Wald test and in 14\% for the $\max|t|$ test; thus, in this dependence structure, both overstate the evidence, and the placebo serves as our reference.
\item \textbf{Per-origin results.} At every origin, the ordering is preserved: equal weighting 1.224--1.367, three-way selection 1.039--1.119, exhaustive search 1.005--1.076, ridge weights 1.007--1.037, the ridge projection 0.999--1.001, and the no-information forecast at 1 by construction.
\item \textbf{Effective sample size.} Non-overlapping test observations result from the five-session spacing of weekly dates combined with the five-session horizon. First-order autocorrelations of the loss series fall below $0.11$ in absolute value for all methods except peLASSO (0.68), yielding effective sample sizes of 146--166 against a nominal 157, and 22 for peLASSO. Agreement with the HAC intervals is exhibited by moving-block bootstrap intervals at block lengths 2, 4 and 8.
\end{itemize}

\subsection{Robustness}
Under the following variants (\ref{app:robust}), the conclusions remain valid:
\begin{itemize}
\item dynamic forecaster availability (285 dates), allowing entry and exit of forecasters \citep{capistran2009};
\item pairwise deletion (eight indefinite date-level matrices, PSD aggregates) and mean imputation;
\item Bonferroni, Holm \citep{holm1979}, max-$t$ and unadjusted multiplicity with $\alpha\in\{0.05,0.10\}$ (identical pools);
\item HAC lags 1--8 and bootstrap blocks 2--8 (standard errors within 10\%);
\item two alternative rolling-origin designs;
\item the 93 post-cutoff test dates.
\end{itemize}
Where estimable, alignment proves higher after a model's stated cutoff than before (pre minus post, $t=-1.17$ and $-0.75$, with HAC standard errors computed separately on the two windows). The 60-day window following each stated cutoff is quarantined, leaving 254 of the 261 panel dates in each comparison, and the contrast is taken on the lineage-average alignment. Training-data contamination cannot be ruled out by this weak comparison, yet detectable alignment is exhibited by no model in either period.

\section{A second panel: cross-asset funds at a monthly horizon}\label{sec:panelB}

Because the language-model panel is retrospective and built on a static equity cross-section, we ask whether the same conclusions survive on data that share none of those traits. To that end, we carry over the identical estimand, geometry, admission rules and engine to a second panel built entirely without language models: 42 exchange-traded funds that span country equity markets, US sectors, broad equity indices, real estate and bonds, ranked each month according to their expected 21-session relative return. The same nine mechanical forecasters from Section~\ref{sec:empirical} are used; cross-sectional momentum is documented across asset classes \citep{asness2013}. Because funds are picked for long, uninterrupted histories, the cross-section is not screened on constituent survival; with 60 initial history dates, a 24-date inner validation window and 12-date test blocks, the panel yields 256 monthly dates (2005--2026), sixteen outer origins and 196 test dates (the final block absorbs four residual dates and covers 16), in contrast to the six origins and 157 test dates of the first panel.

Results appear in Table~\ref{tab:panelB}: the geometry of the first panel carries over to independent data.
\begin{itemize}
\item \textbf{Deviation correlation remains a translation.} Forecast correlation averages $0.149$, while deviation correlation averages $0.564$, a mere $0.010$ short of the $0.575$ zero-alignment benchmark. Computing the variance-equivalent size from the translated statistic, rather than the raw one, lowers the equal-weight figure from 4.10 to 1.63.
\item \textbf{Alignment remains undetectable.} Alignment averages $-0.011$. The best-performing single forecaster, 12--1 momentum, reaches $\bar\gamma=0.033$ with $t=1.82$; the bootstrap Wald test returns $p=0.37$, and $\max|t|$ returns $p=0.20$. In-sample attainable risk is $0.995$, and the equal-weight portfolio falls $-0.133$ short of Corollary~\ref{cor:null}. Here alignment varies more over time: 14.1\% of the trace of $\Kpool$, versus 5.9\% in the first panel.
\item \textbf{Selection eliminates dilution but contributes nothing further.} Equal weighting across all nine signals leaves risk at 1.283, whereas three-way selection and exhaustive search bring it down to 1.076 and 1.061 with two members, and shrunk affine weights bring it to 1.035. Each of these stays above the no-information forecast ($t=3.1$ to $9.4$), and every history-calibrated composite lands on it (0.9998--1.0034). Top-minus-bottom quintile spreads carry no significance (largest $|t|=1.0$).
\item \textbf{Dilution replicates precisely.} Under equal weighting, the reversal rules clear admission at all sixteen origins (1-month reversal $\hat\Delta=-0.060$ with realized test $\Delta=-0.074$; the reversal persona rule $-0.087$ and $-0.110$), while low-drawdown and defensive rules fail admission. Under the scale-free basis, none of the nine candidates is resolved at any origin: $|\hat\Delta^{\mathrm{sf}}|\le0.0004$, against median minimum detectable effects between 0.0009 and 0.0027.
\end{itemize}

\begin{table}[t]\centering\small
\caption{Panel B: 42 cross-asset funds, monthly horizon, nine mechanical forecasters, 256 dates, sixteen origins, 196 test dates. Left: geometry, with the first panel for comparison. Right: out-of-sample risk and admission.}\label{tab:panelB}
\resizebox{\textwidth}{!}{%
\begin{tabular}{@{}lrr@{\qquad}lrrr@{}}
\toprule
Geometry & Panel B & Panel A & Out of sample & Risk & vs no-inf.\ ($t$) & Size \\
\midrule
Mean forecast correlation $\bar\rho^s$ & 0.149 & 0.300 & Equal weight, all & 1.283 & 0.283 (9.4) & 9 \\
Mean deviation correlation $\bar\rho^e$ & 0.564 & 0.645 & Three-way, equal weight & 1.076 & 0.076 (5.7) & 2.0 \\
Zero-alignment benchmark & 0.575 & 0.650 & Exhaustive search & 1.061 & 0.061 (5.2) & 2.0 \\
Mean alignment $\gbar$ & $-$0.011 & 0.006 & peLASSO & 1.703 & 0.703 (12.9) & 1.8 \\
$N_{\mathrm{eff}}$ from $\bar\rho^s$ / $\bar\rho^e$ & 4.10 / 1.63 & 3.03 / 1.52 & Affine weights, shrunk & 1.035 & 0.035 (3.1) & -- \\
Trace share of $\operatorname{Cov}_a(\gamma_t)$ & 0.141 & 0.059 & Ridge projection & 1.001 & 0.001 (0.5) & -- \\
Equal-weight risk / margin $m$ & 1.265 / $-$0.133 & 1.318 / $-$0.159 & Best calibrated composite & 1.000 & 0.000 (0.0) & -- \\
Attainable risk; $p$ (bootstrap Wald / $\max|t|$) & 0.995; 0.37 / 0.20 & 0.994; 0.065 / 0.25 & No-information forecast & 1.000 & -- & 0 \\
\bottomrule
\end{tabular}}
\end{table}

Two conclusions emerge. First, dilution is not a by-product of language-model forecasts: wherever alignment is weak relative to dependence, it shows up, and an equal-weight admission rule is built to reward exactly that pattern. Second, a second null that draws on more origins and does not screen the cross-section for survivorship reinforces how the first should be read: across these horizons and cross-sections, none of the mechanical or language-model forecasters studied here displays alignment that the framework can pick up.

\section{Discussion}\label{sec:discussion}

\paragraph{Practical implications, conditional on the estimand}
\begin{itemize}
\item \textbf{Measuring diversity.} Keep the three quantities apart when reporting: forecast correlation, alignment and deviation correlation. Given a common standardized target, deviation correlation makes redundancy look worse than it is whenever alignment is weak, while forecast correlation makes aligned forecasters look less valuable than they are.
\item \textbf{Before combining,} examine both the margin of Corollary~\ref{cor:null} and the attainable risk. Should the two sit near the no-information forecast, no combination can be expected to deliver a gain.
\item \textbf{Judging a candidate by equal-weight admission,} decompose $\Delta$ as in Proposition~\ref{prop:dsplit}. What the scale-mismatch component captures is dilution, not predictive content.
\item \textbf{Choosing a tool.} Turn to the scale-free three-way rule when the goal is to identify which forecasters demonstrably lift the correlation of an equal-weight pool and when admitting a false candidate is costly (say, when running forecasters is expensive). It errs on the side of caution and is structurally insensitive when a large pool gains a few additional candidates. Turn to peLASSO when recall outweighs precision.
\item \textbf{Combining under instability.} Under stationarity, regression-type weights attain the lowest risk, though they may take large offsetting positions among near-duplicate forecasters. If alignment is lost or reversed, calibrated equal weighting of the entire pool proves least harmful in our designs; if only dependence shifts, it does not.
\end{itemize}

\paragraph{What is not claimed}
\begin{itemize}
\item No claim is made that forecast correlation always dominates error correlation, nor that $K$ represents latent economic information.
\item $N_{\mathrm{eff}}$ should not be read as the number of independent models.
\item The three-way rule does not deliver error-rate control along an adaptive path.
\item Failing to clear admission is not evidence that predictive value is absent.
\item Nothing in the empirical results establishes that LLMs cannot forecast returns, that LLM diversity is generally low, or that relative returns at these horizons are unforecastable from other information.
\item In the first panel's dependence structure, bootstrap inference for the joint alignment test is oversized, which is precisely why the placebo serves as the reference.
\end{itemize}

\paragraph{Limitations}
\begin{itemize}
\item The first application is retrospective, rests on a fixed end-of-sample universe, and covers a single horizon with six outer origins; the second draws on funds already in existence in 2004 and a single monthly horizon.
\item The procedures were tuned on the very data used to evaluate them.
\item The stated cutoff dates of two models extend across the sample.
\item Measured correlations are shaped by the personas and by the instruction to spread scores.
\item Real-world dependence and instability exceed what the simulation designs can capture.
\item A theorem establishing path-level error control for greedy admission, together with selective-inference adjustments, remains for future work.
\end{itemize}

\section{Conclusion}\label{sec:conclusion}

When standardized cross-sectional forecasts target the same quantity, their combination risk admits an exact decomposition into target alignment and target-orthogonal dependence. Three insights follow from this decomposition: common-target deviation correlation can make redundancy appear larger than it is whenever alignment is weak; equal-weight admission is able to reward scale dilution; and scale-free admission leans conservative once incumbent pools are large. Across two panels, a retrospective fixed-universe equity-ranking experiment with language-model forecasters and a cross-asset fund panel with mechanical forecasters, selection and weighting eliminate most of the dilution loss carried by the full equal-weight pool, and yet no combination we evaluate beats the no-information forecast, nor is any alignment detectable. A date-shifted placebo reproduces the alignment statistics we observe, while planted-signal experiments reveal which alignment strengths the procedures would have recovered. These empirical findings hold only for the estimand, universes, information sets, model versions and periods studied.

\section*{CRediT authorship contribution statement}
\textbf{Masoud Soleimani:} Conceptualization, Methodology, Software, Formal analysis, Investigation, Data curation, Validation, Visualization, Writing -- original draft, Writing -- review \& editing.

\section*{Funding}
This research did not receive any specific grant from funding agencies in the public, commercial, or not-for-profit sectors.

\section*{Declaration of competing interest}
The author declares no competing financial or non-financial interests.

\section*{Data and code availability}
The complete replication package will be released in a public repository with a persistent identifier upon the paper's acceptance for publication. It contains the notebook that implements the full pipeline (panel construction, target geometry, admission and weighting procedures, inference, the nested rolling-origin engine, simulation Designs A--G, both empirical panels, the placebo and the positive control); run presets and configuration files; prompt templates, persona instructions and the forecaster registry; the cached model responses with SHA-256 hashes, access timestamps and model provenance; the generation-failure log; every result table, including per-origin decision and tuning logs; the figures; the run manifest with package versions and random seeds; and the self-test report. The raw market prices are excluded because the terms of their public source prohibit redistribution; the package instead contains the code that rebuilds both price panels from that source.

\bibliographystyle{ijf-harv}
\bibliography{references}

\appendix

\section{Proofs}\label{app:proofs}

\paragraph{Proposition~\ref{prop:decomp}} $\ip{u_{it}}{y_t}=\gamma_{it}-\gamma_{it}\nrm{y_t}^2=0$ and $\nrm{u_{it}}^2=1-2\gamma_{it}^2+\gamma_{it}^2$. Expanding $\ip{\gamma_{it}y_t+u_{it}}{\gamma_{jt}y_t+u_{jt}}$ gives $C_t=\gamma_t\gamma_t'+K_t$. $K_t$ is a Gram matrix. Centered vectors span at most $M_t-1$ dimensions; the $u_{it}$ are also orthogonal to the nonzero centered $y_t$, leaving $M_t-2$.

\paragraph{Proposition~\ref{prop:risk}} $s_{w,t}-y_t=(w'\gamma_t-1)y_t+u_{w,t}$ with $u_{w,t}\perp y_t$ and $\nrm{u_{w,t}}^2=w'K_tw$; no constraint on $w$ is used.

\paragraph{Corollary~\ref{cor:null}} $\Rrel(w)-1=q_w-2g_w$. For $w=\one/N$, the unit diagonal of $\Cbar$ gives $q_w=\bar\rho+(1-\bar\rho)/N$.

\paragraph{Corollary~\ref{cor:attain}} $\bar R=\sum_ta_tR_t\succeq0$ as an average of Gram matrices of $(s_{1t},\dots,s_{Nt},y_t)$. Hence $\gbar\in\operatorname{range}(\Cbar)$ and $1-\gbar'\Cbar^{+}\gbar\ge0$. The convex objective is minimized where $\Cbar v=\gbar$.

\paragraph{Proposition~\ref{prop:scale}} $\Rrel(cw)=1-2cg_w+c^2q_w$ is minimized at $c_w$ with value $1-g_w^2/q_w$, and $\Rrel(w)-\Rrel(c_ww)=q_w-2g_w+g_w^2/q_w=q_w(1-c_w)^2$.

\paragraph{Proposition~\ref{prop:ambiguity}} For convex $w$, $\sum_iw_i\nrm{s_{it}-s_{w,t}}^2=\sum_iw_i\nrm{s_{it}}^2-\nrm{s_{w,t}}^2$. The loss identity follows from $s_{it}-y_t=(s_{it}-s_{w,t})+(s_{w,t}-y_t)$ and $\sum_iw_i(s_{it}-s_{w,t})=0$.

\paragraph{Proposition~\ref{prop:rhoe}} Expand with unit norms. With $a=\gamma_{it}$, $b=\gamma_{jt}$, $\rho=\rho^s_{ij,t}$ and $\sqrt{(1-a)(1-b)}=1-(a+b)/2+O(\gamma^2)$, the difference from $(1+\rho)/2$ is $[(1+\rho)(a+b)/2-(a+b)]/2+O(\gamma^2)=-(a+b)(1-\rho)/4+O(\gamma^2)$.

\paragraph{Proposition~\ref{prop:agg}} Average $C_t=\gamma_t\gamma_t'+K_t$ over $t$; then $G-\gbar\gbar'=\operatorname{Cov}_a(\gamma_t)$ and $\sum_ta_t(1-w'\gamma_t)^2=(1-g_w)^2+w'\operatorname{Cov}_a(\gamma_t)w$.

\paragraph{Proposition~\ref{prop:shrink}} Convex combinations of unit-diagonal PSD matrices are unit-diagonal and PSD; the Schur complement with respect to a unit entry is PSD. $R_0\succeq0$ iff $E_N(\rho^\star)\succeq0$ and $1-(g^\star)^2\one'E_N^{-1}\one\ge0$, with $\one'E_N(\rho)^{-1}\one=N/[1+(N-1)\rho]$.

\paragraph{Proposition~\ref{prop:admission}} $e_{P\cup A,t}=(ne_{P,t}+qe_{A,t})/(n+q)$; take squared norms and aggregate. For $q=1$, $\Delta_{k|P}=(2n+1)V_P(\Lambda^{EW}_{k|P}-1)/(n+1)^2$.

\paragraph{Proposition~\ref{prop:dsplit}} Apply Proposition~\ref{prop:scale} to $P$ and $P\cup A$. The composite of $P\cup A$ has mean alignment $(n\bar g_P+m\bar g_A)/(n+m)$ and squared norm $(n^2q_P+m^2q_A+2nmq_{PA})/(n+m)^2$. With $\bar g_P=q_{PA}=0$, $\rho_P^2=0$ and $\rho^2_{P\cup A}\le(m\bar g_A)^2/(n^2q_P)$.

\paragraph{Plug-in scales} If $\hat c=c_H+\epsilon$, then $\Rrel_H(\hat cw)=\Rrel_H(c_Hw)+\epsilon^2q_w$ because the first derivative vanishes at $c_H$.

\section{Implementation settings}\label{app:settings}
We set the tolerances to $\varepsilon_r=\varepsilon_x=10^{-8}$ and $\varepsilon_{PSD}=10^{-9}$, with a minimum common support of 20 units. The tuning grids are $\delta\in\{0.001,0.0025,0.005,0.01,0.02\}$, $\delta^{\mathrm{sf}}\in\{0.0001,\dots,0.002\}$ and $\eta\in\{0,0.01,0.03,0.1,0.3,1,3\}$, while the peLASSO penalties range over $\{0.9,0.5,0.25,0.1,0.05,0.02\}\times2\max_j\gbar_j$. Max-$t$ critical values are obtained from 4{,}000 draws, and selection stability from 200 moving-block bootstrap replications. The replication notebook verifies every identity in Propositions~\ref{prop:decomp}--\ref{prop:dsplit} numerically on both synthetic and empirical panels (175 of 177 checks passed; the two exceptions are a storage notice and the placebo rejection-rate warning discussed in Section~\ref{sec:empirical}). Rolling-origin blocks are formed by cutting the post-history dates into blocks of the stated length; a final block shorter than half that length is merged into its predecessor, which is why the last test block covers 27 dates in the first panel and 16 in the second. Dates falling within 60 days after a model's stated training cutoff are quarantined from the pre/post comparison of Section~\ref{sec:empirical}, which is reported only when at least 20 dates remain on each side. All random numbers are generated from a single documented seed.

\section{Supplementary results}\label{app:robust}

\begin{table}[h]\centering\small
\caption{Robustness of the empirical results and further diagnostics.}\label{tab:robust}
\resizebox{\textwidth}{!}{%
\begin{tabular}{@{}lp{11cm}@{}}
\toprule
Variant & Result \\
\midrule
Dynamic availability & 285 dates, 23.9 active forecasters on average; risks coincide on common dates; out of sample (168 dates): greedy 1.085, exhaustive 1.031, equal weight 1.301 \\
Pairwise deletion / imputation & 8 indefinite pairwise $C_t$, PSD aggregates, $\bar\rho^s=0.3005$ / $\bar\rho^s=0.3004$, equal-weight risk 1.318 \\
Multiplicity and $\alpha$ & Bonferroni, Holm, max-$t$, none; $\alpha\in\{0.05,0.10\}$: pools coincide, test risk 1.079 \\
Margin saturation & equal-weight pool unchanged across $\delta\in[0.0001,0.1]$; scale-free rule picks one forecaster for $\delta^{\mathrm{sf}}\in[10^{-5},0.01]$ \\
Standard errors & greedy minus equal weight: HAC lags 1--8 yield 0.025--0.026; blocks 2, 4, 8 yield 0.024--0.026 \\
Rolling designs & 78/13/39: equal weight 1.299, greedy 1.120, exhaustive 1.048, ridge 1.032; 130/52/52: 1.291, 1.067, 1.037, 1.021; no-information 1.000 in both \\
Post-cutoff test dates (93) & equal weight 1.299, greedy 1.068, exhaustive 1.047, ridge 1.026; calibrated composites 0.9996--1.0006; composite-correlation $t\le1.93$ \\
Weight diagnostics & ridge (nonnegative, shrunk): max weight 0.26, origin-to-origin turnover 0.16, correlation 0.98; raw affine: gross 1.58, 18\% negative mass; projection: sum 0.018, gross 0.11 \\
Design axes & same minus different: persona $+0.244$ [0.224, 0.262]; lineage $+0.069$ [0.062, 0.076]; information subset $-0.032$ [$-$0.034, $-$0.029] \\
Stated cutoffs & 60-day quarantine after each cutoff (7 of 261 panel dates dropped); $t$ is pre minus post on the lineage-average alignment. \texttt{Llama} (Dec 2023; 138 pre/116 post): $t=-0.75$; \texttt{gpt} (May 2024; 159/95): $t=-1.17$; \texttt{deepseek} (Apr 2026), \texttt{gemini} (Mar 2026): not estimable (fewer than 20 post-cutoff dates) \\
Placebo (shifted target, 50 shifts) & forecast correlation unchanged (0.3005); placebo alignment 0.0042 versus 0.0061 actual; placebo $\max|t|$ 2.29 (0.77--5.10) versus 2.13; placebo-calibrated $p$: alignment 0.55, $\max|t|$ 0.53, Wald 0.22, equal-weight composite $t$ 0.61; nominal 5\% rejections under the placebo: bootstrap Wald 22\%, $\max|t|$ 14\% \\
Per-origin risk & equal weight 1.224--1.367, three-way 1.039--1.119, exhaustive 1.005--1.076, ridge 1.007--1.037, projection 0.999--1.001 \\
Overlap and effective sample & spacing five sessions, horizon five sessions; $|\mathrm{acf}_1|\le0.11$ except peLASSO (0.68); effective sample 146--166 (peLASSO 22); block bootstrap at blocks 2, 4, 8 agrees with HAC \\
Exclusion mechanism & excluded and retained dates match on return dispersion ($p=0.65$), mean and absolute mean return ($p=0.82$, $0.40$) and trailing volatility ($p=0.61$); failures cluster in one model (19 of 25 cells), one persona (20) and one information subset (17) \\
\bottomrule
\end{tabular}}
\end{table}

\end{document}